\documentclass[aps,prl,twocolumn,superscriptaddress,showpacs]{revtex4}
\usepackage{amssymb}
\usepackage{amsmath}
\usepackage{graphicx}

\newcommand{\reffig}[1]{Fig.~\ref{#1}}

\begin{document}

\title{Ultrafast spatio-temporal dynamics of terahertz generation by
ionizing two-color femtosecond pulses in gases}
\author{I. Babushkin}
\affiliation{Weierstra\ss -Institut f\"ur Angewandte Analysis und Stochastik, 10117
Berlin, Germany}
\author{W. Kuehn}
\affiliation{Max-Born-Institut f\"ur Nichtlineare Optik und Kurzzeitspektroskopie, 12489
Berlin, Germany}
\author{C. K\"ohler}
\affiliation{Max Planck Institute for the Physics of Complex Systems, 01187 Dresden,
Germany}
\author{S. Skupin}
\affiliation{Max Planck Institute for the Physics of Complex Systems, 01187 Dresden,
Germany}
\affiliation{Friedrich Schiller University, Institute of Condensed Matter Theory and
Optics, 07742 Jena, Germany}
\author{L. Berg\'e}
\affiliation{CEA-DAM, DIF, F-91297 Arpajon, France}
\author{K. Reimann}
\affiliation{Max-Born-Institut f\"ur Nichtlineare Optik und Kurzzeitspektroskopie, 12489
Berlin, Germany}
\author{M. Woerner}
\affiliation{Max-Born-Institut f\"ur Nichtlineare Optik und Kurzzeitspektroskopie, 12489
Berlin, Germany}
\author{J. Herrmann}
\affiliation{Max-Born-Institut f\"ur Nichtlineare Optik und Kurzzeitspektroskopie, 12489
Berlin, Germany}
\author{T. Elsaesser}
\affiliation{Max-Born-Institut f\"ur Nichtlineare Optik und Kurzzeitspektroskopie, 12489
Berlin, Germany}

\begin{abstract}
We present a combined theoretical and experimental study of  spatio-temporal
propagation effects in terahertz (THz) generation in  gases using two-color
ionizing laser pulses. The observed strong  broadening of the THz spectra
with increasing gas pressure reveals  the prominent role of spatio-temporal
reshaping and of a  plasma-induced blue-shift of the pump pulses in the
generation  process. Results obtained from (3+1)-dimensional simulations are
in  good agreement with experimental findings and clarify the mechanisms
responsible for THz emission.
\end{abstract}

\maketitle


Far-infrared radiation in the THz range has developed into a sensitive probe
of low-frequency excitations of condensed matter and into an analytical and
imaging tool for a broad range of applications. Recently, intense THz pulses
have been applied for inducing nonlinear light-matter interactions and for
studying quantum-coherent charge transport phenomena in solids. In this new
area of research, the generation of well-defined THz field transients with a
high amplitude represents a key issue. Both electron accelerator- and
laser-based sources have been developed to generate THz transients.
Conventional laser-driven THz sources are based on semiconductor
photoconductive switches \cite{1} and nonlinear frequency conversion in
crystals \cite{2}, providing a comparably small THz field strength in a
spectral range limited by the absorption of the crystals. In an alternative
approach \cite{3}, a short pump pulse at 800~nm and its second harmonic at
400~nm are focused into a gaseous medium to generate a plasma. Intense THz
pulses with field amplitudes as high as 400~kV/cm and remarkably broad
spectra have been reported with this method \cite{3,4,5,6,7,8,9,10,11,12,13}.
However, the physical mechanisms behind the observed THz generation
remain controversial. Initially, THz generation has been explained by
four-wave mixing rectification \cite{3,penano10} and experimental
results have been interpreted along those lines (see, e.g.,
\cite{4,5,6,8,13}). In contrast, other studies attribute THz emission
to a laser-induced plasma current in the asymmetric two-color laser
field \cite{9,10,12}. So far, theoretical work has mainly focused on
the analysis of local fields by using photocurrent models. An
alternative approach is based on particle in cell simulations
\cite{16,17,19} which are computationally very expensive and limited
to propagation lengths of a few micrometers only.

In this Letter we present a combined theoretical and experimental study of
THz generation by ionizing two-color femtosecond pulses in a gas. Extensive
numerical simulations were performed using for the first time a
(3+1)-dimensional code based on an unidirectional pulse propagation equation
which includes the plasma dynamics responsible for the observed THz
generation. This approach allows a comprehensive description of the
propagation of all fields taking into account their spatio-temporal
reshaping induced by the plasma effects and the optical Kerr nonlinearity.
We demonstrate that spatio-temporal propagation effects are indispensable
for understanding the generation process and influence the THz spectrum
substantially. In both experiments and simulations we observe a remarkable
broadening of the THz spectrum with increasing gas pressure. Such broadening
is a result of a sensitive dependence of the THz spectrum on small phase and
frequency shifts induced by nonlinear propagation of the fundamental  and
second harmonic pulses. We show that the stepwise character of the ionization
process on
the sub-femtosecond time scale is essential for such dependence.

\begin{figure}[tbp]
\includegraphics[width=8.5cm]{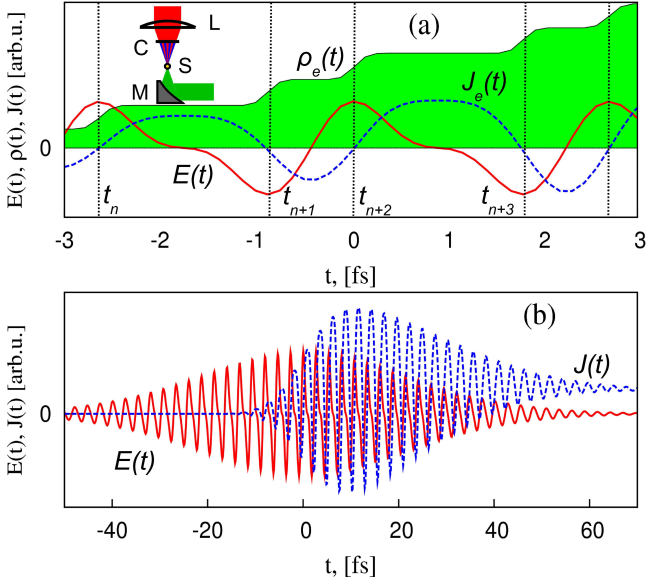}
\caption{(color online) The mechanism of THz generation: (a) The
  two-color electric field $E$ generates free electrons with a
  step-wise modulation of the electron density $\protect\rho$ via
  tunneling photoionization. The ionization occurs mostly near the
  maxima of electric field at time instances $t_n$.  (b) This leads to
  a slow component of the current $J$ growing on time scales of the
  pulse duration, which acts as a source for THz emission. Inset in
  (a): scheme of the experimental setup.} \label{setup}
\end{figure}

Our THz plasma source (inset of Fig.~\ref{setup}) is driven by $40$~fs
pulses (800~nm) with pulse energies of $\sim 300~\mu $J at a repetition
rate of $1$~kHz from a Ti:sapphire laser system. Intensities far above the
field ionization threshold are reached by focusing the beam of 8 mm diameter
with an achromatic lens (L) of $40$~mm focal length. A $0.1$~mm thin
$\beta$-barium 
borate (BBO) crystal (C) cut for type~I second-harmonic generation
is additionally inserted into the convergent beam $7$~mm before the focus.
The setup is placed in a closed chamber filled with argon at various
pressures between 1 and 1000~mbar. THz radiation emitted from the plasma
volume in the focal spot is collected by a parabolic mirror (M) with a
diameter of $25.4$~mm at a distance of its effective focal length of $12.7$~mm.
The THz spectrum is measured by a Michelson interferometer. Intensity
interferograms were recorded with a mercury cadmium telluride (HgCdTe)
detector by varying the path difference between the two  arms, and the THz
spectra were obtained by Fourier transformation.

For the description of the nonlinear propagation of electromagnetic fields
valid in the full range from THz down to UV/VUV wavelengths beyond paraxial
and slowly-varying envelope approximations we use the following
unidirectional pulse propagation equation \cite{Kolesik:pre:70:036604} for
linearly polarized pulses 
\begin{equation}
\partial _{z}\hat{E}=i\sqrt{k(\omega )^{2}-k_{x}^{2}-k_{y}^{2}}\hat{E}+i%
\frac{\mu _{0}\omega ^{2}}{2k(\omega )}\hat{\mathcal{P}}_{\mathrm{NL}}.
\label{Eq.(1)}
\end{equation}%
Here, $\hat{E}(k_{x},k_{y},z,\omega )$ is the Fourier transform (indicated
by $\hat{~}~$) of the electric field with respect to $x$, $y$, and $t$, $%
k(\omega )=\omega n(\omega )/c$ is the wavenumber, $\omega $ the
frequency, $c$ 
is the speed of light and $n(\omega )$ is the refractive index of the
gaseous medium, in our case argon \cite{dalgarno}. The first term on the
r.h.s.\ of Eq.~(\ref{Eq.(1)}) describes {linear} dispersion and diffraction
of the pulse. The nonlinear polarization
$\hat{\mathcal{P}}_{\mathrm{NL}}=
\hat{P}_{\mathrm{Kerr}}+i\hat{J}_{e}/\omega +i\hat{J}_{\mathrm{loss}}/\omega
$ originates from the optical Kerr effect $P_{\mathrm{Kerr}}$, the electron
current ${J_{e}}$ and a loss term $J_{\mathrm{loss}}$ due to photon
absorption during ionization. In $\hat{P}_{\mathrm{Kerr}}$ we take into
account
different nonlinear susceptibilities for neutral atoms and ions
\cite{20}.
However, the Kerr response has negligible influence on the THz spectrum.
The plasma dynamics is described by the electron density $\rho _{e}(t)$,
obeying 
\begin{equation}
\dot{\rho}_{e}(t)=W_{\mathrm{ST}}(E)[\rho _{at}-\rho _{e}(t)],
\label{Eq.(3a)}
\end{equation}%
where $\rho _{at}$ denotes the neutral atomic density and dot the
time-derivative. We use the quasi-static tunneling ionization rate for
hydrogen-like atoms given by \cite{10} $W_{\mathrm{ST}}(E)=4\omega
_{a}(r_{H})^{2.5}[E_{a}/|E|]\exp [-2(r_{H})^{1.5}E_{a}/3|E|]$, where 
$E_{a}=m_{e}^{2}q^{5}/\hbar ^{5},\omega _{a}=m_{e}q^{4}/\hbar ^{3}$
and 
$r_{H}=U_{h}/U_{Ar}$. $U_{h}$ and $U_{Ar}$ are the ionization potentials of
hydrogen and argon; 
$m_{e}$ and $q$ are the electron mass and charge, respectively. 
For the anticipated electric field strengths, we neglect multi-photon and
avalanche ionization. The transverse macroscopic plasma current $J_{e}(t)$
is determined by the microscopic velocity distribution $v(t,t_{0})$ of
electrons born at the time $t_{0}$ \cite{9,10,12}, 
\begin{equation}
J_{e}(t)=q\int\limits_{-\infty }^{t}v(t,t_{0})\dot{\rho _{e}}(t_{0})dt_{0}.
\label{eq:jmic}
\end{equation}%
Assuming zero velocity for new-born electrons
and neglecting the influence of the magnetic field and electron-electron
interaction, the electron velocity reads $v(t,t_{0})=\frac{q}{m_{e}}%
\int_{t_{0}}^{t}E(\tau )\exp [-\nu _{e}(t-\tau )]d\tau $, where $\nu _{e}$
is the electron-ion collision rate. 
Substituting this into Eq.~(\ref{eq:jmic}) we obtain 
\begin{equation}
\dot{J}_{e}(t)+\nu _{e}J_{e}(t)=\frac{q^{2}}{m_{e}}E(t)\rho _{e}(t).
\label{Eq.(6)}
\end{equation}%
The above analysis demonstrates that we regain a well-known equation for the
current (see, e.g., \cite{20}).
To formally ensure
energy conservation during ionization, the additional term ${J}_{\mathrm{loss%
}}=W_{\mathrm{ST}}[E](\rho _{\mathrm{at}}-\rho _{e})U_{Ar}/E
$ is introduced in Eq.~(\ref{Eq.(1)}) such that the energy dissipation $J_{%
\mathrm{loss}}E$ equals the ionization energy loss.


Let us now illustrate the mechanism behind THz generation. In
Fig.~\ref{setup}(a) we present the electron density (in green), which
shows a stepwise increase near the tunnel ionization events at the
field maxima (red curve).  Such behavior has been recently observed in
real-time experiments with sub-femtosecond resolution
\cite{Uiberacker}.  In a simplified model, we can assume rectangular
steps in the electron density [$\dot{\rho}%
_{e}(t)=\Sigma _{n}\rho _{n}\delta (t-t_{n})$] and $\nu_e=0$, and
thereby obtain a discrete version of Eq.~(\ref{eq:jmic}):
\begin{equation}
J_{e}(t)\sim \sum_{n}\rho _{n}H(t-t_{n})[v_{f}(t)-v_{f}(t_{n})],
\label{model}
\end{equation}
 where $H(t)
$ is the Heaviside step function, $v_{f}(t)=\frac{q}{m_{e}}\int_{-\infty
}^{t}E(\tau )d\tau
$ is the free electron velocity such that
$v(t,t_{n})=[v_{f}(t)-v_{f}(t_{n})]$, 
and $\rho _{n}$ is the electron density created in the $n$th ionization
event. For a monochromatic
electric field $v_{f}(t_{n})=0$. 
The Fourier transformation of the step function $H(t)$ exhibits a low-frequency  spectrum $\sim 1/\omega$,
therefore THz radiation is generated by the terms in Eq.~(\ref{model})
proportional to $v_{f}(t_{n})$, while the 
terms proportional to $v_{f}(t)$
contribute in the spectral range of the pump fields. 

\begin{figure}[tbp]
\includegraphics[width=8.5cm]{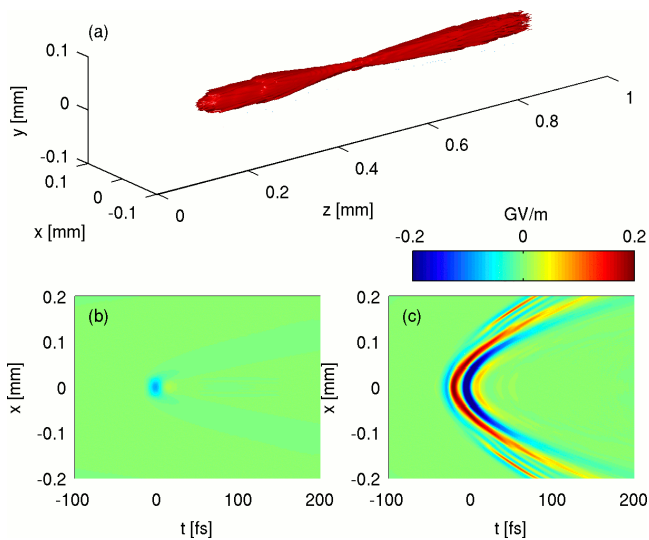}
\caption{(color online) (a) Iso-electron-density surface at $\protect\rho%
_e=5\times10^{17}$~cm$^{-3}$ for 200~mbar gas pressure. Computed THz fields (%
$\protect\omega<500$~ps$^{-1}$) at (b) $z=0.2$~mm and (c) $z=1$~mm.}
\label{3d}
\end{figure}

In order to model the experimental conditions in the plasma spot, we
consider a Gaussian input beam with a waist $w_{0}=100~\mu $m and a duration
of the 800~nm pump pulse of (FWHM) $t_{p}=40$~fs (pulse energy 300 $\mu$J).
The energy of the second harmonic at 400~nm is chosen as 12\% of the
fundamental as estimated from the experiment. The duration $t_{p}$ and the
waist $w_{0}$ of the second harmonic are by a factor $\sqrt{2}$ smaller than
the values for the fundamental. The phase angle between the two components
is zero initially and shifts during propagation to non-zero values. The
pulsed input beams are focused ($f=0.5$~mm, in order to have a comparable
ratio $w_{0}/f$ in experiment and simulation) into the argon atmosphere.
Figure~\ref{3d} shows (a) the evolution of the computed plasma density and
the transverse distribution of the THz field at (b) 0.2~mm and (c) 1~mm for
a gas pressure of 200~mbar. The field intensities reach the ionization
threshold shortly after the starting point of the simulation at $z=0$~mm,
and form a 0.7~mm long plasma channel. THz fields inside this focal region
reach values of the order of GV/m and exhibit a strong diffraction.

\begin{figure}[tbp]
\includegraphics[width=8.5cm]{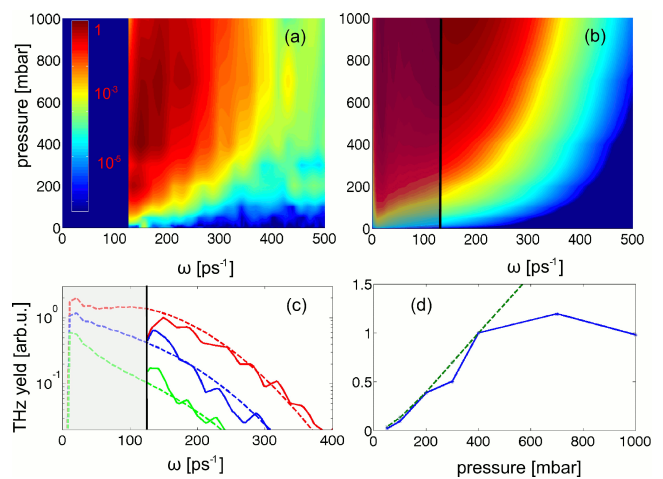}
\caption{(color online). Measured THz spectra (a) and simulation results (b)
for pressures between 1 and $1000$~mbar. In (c), experimental (solid lines)
and theoretical (dashed lines) spectra are compared for $100$, $200$ and $400
$~mbar (green, blue and red curves, resp.). In (b),(c), the part below $%
\protect\omega=125$~ps$^{-1}$ is shaded, because no experimental values are
available. In (d), the overall THz yield versus pressure is shown (dashed
green line represents the simulation, solid line the experiment).}
\label{exp1}
\end{figure}

Let us now investigate the dependency of THz generation on the gas pressure.
Measured spectra over the complete pressure range between $1$~mbar and
$1000$~mbar are shown as a contour plot in Fig.~\ref{exp1}(a). Our HgCdTe
detector
is sensitive in a frequency range $\omega =2\pi \nu $ from $125$~ps$^{-1}$
to $1070$~ps$^{-1}$, i.e., the drop of the measured energy spectral density
around 125~ps$^{-1}$ is due to the decreasing sensitivity. For a comparison
of theory and experiment, the high-frequency part of the spectra is most
relevant. In our setup, we mainly detect THz radiation generated in the
focus of the mirror (M). Simple ray tracing estimates indicate that the
length of this focal region is $\lesssim 0.3$~mm. The almost vanishing
spectrum at small argon pressure shows clearly that the plasma and not the
BBO crystal acts as a source of the emitted radiation. In the region from
zero to $\sim 300$~mbar the spectral width increases strongly. The highest
frequencies even beyond $300$~ps$^{-1}$ are detected at pressures larger
than $300$~mbar. Above $\sim 500$~mbar the slope of the high frequency wing
stays rather constant. The measured THz yield [solid line in Fig.~\ref{exp1}%
(d)] grows linearly with the pressure up to $400$~mbar before it saturates.

In Fig.~\ref{exp1}(b,c), THz spectra computed with our simulation code are
shown. The simulated spectra in Fig.~\ref{exp1}(b,c) are obtained by
integration over the transverse coordinates $(x,y)$. We find very good
agreement between experiment and simulation below 500~mbar for THz fields
generated at the beginning of the plasma spot around $z=0.2$~mm ($0.3$~mm
before the linear focus). THz fields generated upon further propagation
become spectrally much broader. Hence, we conjecture that the parabolic
mirror in the experiment images the leading part of the plasma spot only.
The calculated THz yield increases linearly with gas pressure [dashed line
in Fig.~\ref{exp1}(d)], in agreement with experimental results up to
$400$~mbar. The saturation of the experimental yield at higher pressure is
likely
due to additional THz losses upon further propagation towards the mirror,
whereas the simulated yield is computed directly at the position $z=0.2$~mm.

The observed pressure dependence of the THz spectral maximum and width gives
insight into important features of plasma-induced THz generation. The
dependence of the spectral width on pressure can not be explained by the
local plasma current in which the variation of pressure results only into an
amplitude scaling of the current. Instead, it originates from pressure
dependent nonlinear propagation effects. For the intensity range and plasma
interaction length of the experiment, the calculated spectral evolution of
the pump pulses at 400~nm ($\omega = 2\pi \times 750$ THz) and 800~nm ($%
\omega = 2\pi \times 375 $ THz) shows that their spectral broadening is
negligible. We observe, however, small blue-shifts $\delta \omega$ of the
central frequencies caused by the nonlinear plasma-induced change of the
refraction index \cite{21,22}. These shifts are $\sim 6$~ps$^{-1}$ in the
fundamental and $\sim 2.5$~ps$^{-1}$ in the second harmonic at $z=0.2$~mm for
400~mbar, and depend strongly on the gas pressure.

\begin{figure}[tbp]
\centerline{\includegraphics[width=8.5cm]{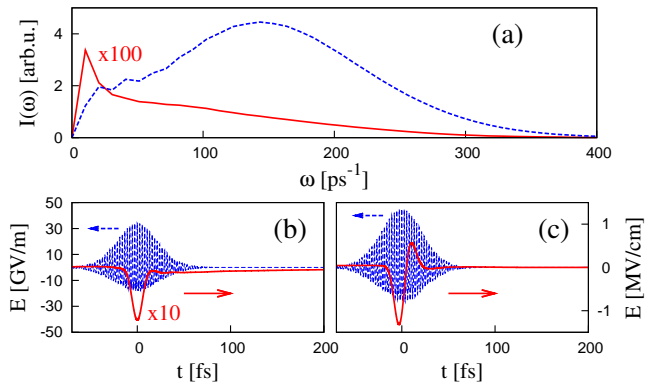}}
\caption{(color online) (a) On-axis $(x=0,y=0)$ spectral intensity $I(%
\protect\omega)$ and (b,c) temporal shape of $E(t)$ for 400~mbar gas
pressure. In (a), $I(\protect\omega)$ at $z=0.1$~mm and $z=0.2$~mm are shown by
the solid red and dashed blue line,
respectively. The temporal profiles of the field $E(t)$ (blue curve) and its
low frequency part (red curve) are shown in (b) for $z=0.1$ mm and in (c)
for $z=0.2$ mm. $I(\omega)$ and $E(t)$ at $z=0.1$~mm [red curves in (a) and (b)]
are amplified by a factor 100 and 10, respectively, in order to improve
visibility. }
\label{fig:thz}
\end{figure}

Surprisingly, these very small frequency shifts have a dramatic influence on
the generated THz spectrum. This effect is most pronounced in the on-axis
spectra, where the intensity is maximal. In Fig.~\ref{fig:thz}(a), the
calculated on-axis THz spectra are plotted for 400~mbar gas pressure,
displaying dramatic changes in spectral shape and the maximum
frequency between 0.1~mm (red curve) and 0.2~mm (blue curve). Figs.
4 (b,c) show the corresponding time-dependent field and its low-frequency part.
To explain the physical origin of this very sensitive dependence, let us go
back to our simplified model [Eq.~(\ref{model})].
For a two-color optical field, $%
E=A_{1}\cos [(\omega _{0}+\delta \omega )t]+A_{2}\cos (2\omega _{0}t+\theta )$,
the field maxima in every half-cycle are given by $\omega
_{0}t_{n}\approx n\pi -n\pi \delta \omega /\omega _{0}-(-1)^{n}2r\sin \theta
$, provided that $A_{2}/A_{1}=r\ll 1$ and $n\delta \omega
\ll \omega _{0}$. 
Hence, the points in time $t_{n}$ and therefore the free velocities
$v_{f}(t_{n})$ alter significantly when $\delta \omega $ (and $\theta$)
change upon the propagation. As seen above, the low-frequency spectrum is determined
by a sum over contributions $\sim v_{f}(t_{n})$, and this sum finally determines
the THz spectral shape in \reffig{fig:thz}(a). In the full
(3+1)-dimensional geometry, the spectral shapes generated at
different spatial points are added and averaged, leading to the
strong spectral broadening observed in Fig.~\ref{exp1}(c).
Thus, the above described dependence of the THz spectral shapes on pressure and
propagation distance can be explained by propagation effects of the pump
fields. We would like to
stress that phase and frequency relation of fundamental and second harmonic
fields always change during propagation.

In conclusion, we presented a theoretical and experimental investigation of
photo-current induced THz generation. Using (3+1)-dimensional simulations
and experimental measurements we show a strong dependence of the THz spectra
on gas pressure. Our results give insight into the important influence of
nonlinear propagation effects and the mechanism of THz generation.
Plasma-induced blue-shifts of the driving pulses play a key role in the
pressure dependent broadening of the THz spectra and confirm that the
emission process of THz radiation is associated with a stepwise modulation
of the tunneling ionization current. We believe that our findings open
interesting perspectives to control THz emission in a broader spectral range.

This work has been performed using HPC resources from GENCI-CINES (Grant
2009-x2009106003).


\begin{thebibliography}{99}
\bibitem{1} D. You, R. R. Jones, P. H. Bucksbaum, and D. R. Dykaar, Opt.
Lett. \textbf{18}, 290 (1993).

\bibitem{2} D. H. Auston, K. P. Cheung, J. A. Valdmanis, and D. A. Kleinman,
Phys. Rev. Lett. \textbf{53}, 1555 (1984).

\bibitem{3} D. J. Cook and R. M. Hochstrasser Opt. Lett. \textbf{25}, 1210
(2000).

\bibitem{4} P. Zhang et al., Phys. Rev. Lett. \textbf{91}, 225001 (2003).

\bibitem{5} M. Kress et al.
Opt. Lett. \textbf{29}, 1120 (2004).

\bibitem{6} T. Bartel et al., Opt. Lett. \textbf{30}, 2805 (2005).

\bibitem{7} M. Kress et al., Nat. Phys. \textbf{2}, 327 (2006).

\bibitem{8} Xu Xie, J. Dai, and X.-C. Zhang, Phys. Rev. Lett. \textbf{96},
075005 (2006).

\bibitem{9} K. Kim, J. Glownia, A. Taylor, and G. Rodriguez, Opt. Expr.
\textbf{15}, 4577 (2007).

\bibitem{10} M. D. Thomson, M. Kress, T. Loeffler, H. G. Roskos, Laser \&
Photon. Rev. \textbf{1}, 349 (2007).

\bibitem{11} K. Reimann, Rep. Progr. Phys. \textbf{70}, 1597 (2007).

\bibitem{12} K. Kim, J. Glownia, A. Taylor, and G. Rodriguez, Nat. Phot.,
\textbf{2}, 605 (2008).

\bibitem{13} A. Houard, Yi Liu, B. Prade and A. Mysyrowicz, Opt. Lett.
\textbf{33}, 1195 (2008).

\bibitem{penano10} J. Pe\~nano et al., Phys. Rev. E \textbf{81} 026407
(2010).

\bibitem{16} H.-Ch. Wu, J. Meyer-ter-Vehn, and Zh.-M. Sheng, New J. Phys.
\textbf{10}, 043001 (2008).

\bibitem{17} M. Chen, A. Pukhov, X.-Yu Peng, and O. Will, Phys. Rev. E
\textbf{78}, 046406 (2008).

\bibitem{19} W.-M. Wang et al., Opt. Expr. \textbf{16}, 16999 (2008).

\bibitem{Uiberacker} M. Uiberacker et al., Nature \textbf{446}, 627 (2007) .

\bibitem{dalgarno} A.~Dalgarno and A.~E.~Kingston, Proc.\ Royal Soc.\ London
A \textbf{259}, 424 (1960).

\bibitem{20} P. Sprangle, J. R. Penano, B. Hafizi, and C. A. Kapetanakos,
Phys. Rev. E \textbf{69}, 066415 (2004).

\bibitem{Kolesik:pre:70:036604} M.~Kolesik and J.~V. Moloney, Phys.\ Rev.\ E
\textbf{70}, 036604 (2004).

\bibitem{21} W. M. Wood, C. W. Siders, and M. C. Downer, Phys. Rev. Lett.
67, 3523 (1991).

\bibitem{22} S.C. Rae and K. Burnett, Phys. Rev. A 46, 1084 (1992).

\end{thebibliography}
\end{document}